# Maximum-Bandwidth Node-Disjoint Paths

Mostafa H. Dahshan
Dept. of Computer Engineering
College of Computer and Information Sciences
King Saud University
Riyadh, Saudi Arabia

*Abstract*—This paper presents a new method for finding the node-disjoint paths with maximum combined bandwidth in communication networks. This problem is an NP-complete problem which can be optimally solved in exponential time using integer linear programming (ILP). The presented method uses a maximum-cost variant of Dijkstra algorithm and a virtual-node representation to obtain the maximum-bandwidth node-disjoint path. Through several simulations, we compare the performance of our method to a modern heuristic technique and to the ILP solution. We show that, in a polynomial execution time, our proposed method produces results that are almost identical to ILP in a significantly lower execution time.

## I. INTRODUCTION

Path optimization is a fundamental problem in data networks. Traditional path optimization aims to find the single lowest-delay path between a given source and destination nodes. The main application for such a problem is routing in IP networks. For other applications, variants of the problem are needed. QoS requirements may set one or more constraints to be satisfied along the path [1]. In general, QoS constraints can be classified into additive, multiplicative, and concave. Additive constraints include: delay, jitter, and hop count. Multiplicative constraints include the probability of packet arrival and link reliability. Concave constraints include finding the minimum or maximum bandwidth along the path to represent the bandwidth of the path [2]. In addition to single path QoS requirements, recovery plans may require having one or more backup paths to be ready in case the primary path fails. Multiple paths are also required in traffic engineering schemes to provide load balancing [3], [4]. Multiple paths usually have an additional constraint to be link-disjoint or node-disjoint. Node-disjoint paths are usually harder to find but provide more robustness in case of node failures.

The complexity of path optimization varies depending on the type and number of constraints. Several studies have shown that the Multiple Constrained Path (MCP) problems are generally NP-complete and are not solvable in polynomial time [5], [6]. Furthermore, finding disjoint paths with a single constraint is generally an NP-complete or NP-hard problem [7], [8].

In this paper, our focus is on the problem of finding two node-disjoint paths such that the bandwidth sum of the two paths is the maximum possible two-disjoint-paths sum between a given source and destination nodes in the network. This is essentially an MCP problem with two constraints: The first constraint is for the two paths to be node-disjoint. The second constraint is maximizing the bandwidth sum of the two paths. This is also an NP-complete problem as shown in [9]. We develop a near-optimal method for solving this problem in polynomial time. The proposed method uses a virtual-node representation from the original network. We implement a variant of Dijkstra algorithm that finds the optimal path based on the maximum bandwidth [10]. The variant algorithm is further modified to work concurrently on two paths, avoiding nodes that lead to overlapped paths in the original network. The algorithm is then applied iteratively to obtain the maximum disjoint path in the actual network.

In the remaining part of the paper, we discuss related studies that attempted to find solutions to the maximum-pair disjoint paths and similar problems. Next, we illustrate the modified Dijkstra algorithm that finds the maximum-bandwidth path. The new method is then presented in details and demonstrated by an example. The performance of our method is evaluated and compared to a modern heuristic algorithm and to the exact solution using ILP. Analytical study of the presented method is then presented to show the order of its execution time. The paper is concluded with a summary and future work.

## II. RELATED WORK

The problem of finding disjoint paths has been subject to intensive research. In particular, several studies have addressed the problem of finding maximum combined bandwidth in disjoint paths. Shen and Sen [9] have discussed two versions of the problem. The first version is finding pair of disjoint paths with maximum combined bandwidth which they call "widest pair of disjoint paths". The second version is finding a pair of disjoint paths such that the bandwidth of the first path is greater than or equal to X1 and the bandwidth of the second path is greater than or equal to X2. They proved that both versions of the problem are NP-complete and provided both exact solutions using ILP and two approximate heuristic solutions.

The first solution, deterministic heuristic algorithm (DHA), works in two phases: First, it uses a relaxed version of ILP, in which solutions are not necessarily integers. If it produces integer values, then the solution is accepted. If not, the second phase replaces the capacity of each edge with values from the solution obtained in the phase 1 and then applies Suurballe's algorithm [11], [12] to attempt finding the two paths. The second solution, randomized heuristic algorithm (RHA), uses the same phase 1 as in DHA. The second phase constructs two graphs g1 and g2 with edges assigned values obtained from ILP in phase 1, first path for g1 and second path for g2. Then, it uses random walks to get the two disjoint paths by alternating the random function between g1 and g2. Despite the relaxation of ILP in these heuristics, the time required for ILP execution is generally in exponential order.

Shen et al [13] have addressed the problem of maximizing bandwidth through disjoint paths. They proposed a heuristic algorithm called Algorithm-1. The algorithm generates all possible paths from source to destination using the algorithm of Am et al [14]. The algorithm then creates a path intersection graph G' in which each node represents one of the paths obtained in step 1. Paths are considered intersected if they share at least a common edge. Nodes of the G' have weights to indicate the bandwidth of the path. Finally, the algorithm finds the maximum weight independent set S of G', which are nodes in G' with maximum weights and which are not connected by edges. Finding maximum independent set is an NP-complete problem and thus approximate algorithms are used.

Leng et al [15] have studied a different but related problem of finding shortest pair of disjoint paths with bandwidth guarantee (SPDP-BG). In this problem, it is required to find a pair of disjoint paths with minimum cost, while guaranteeing a minimum bandwidth of a defined value X. They proved that the SPDP-BG problem is NP-complete and presented a heuristic algorithm to solve the problem. Their heuristic algorithm finds a pair of disjoint paths with guaranteed minimum bandwidth, and then modifies them to gradually minimize their lengths. The algorithm first finds the widest-bandwidth path.

Next, it uses the aforementioned DHA algorithm of [9] to get the widest pair of disjoint paths. The length of the two disjoint paths is used as an upper bound for the path length. After that, find k-shortest paths using the algorithm of [16]. Next, loop in each of the k-shortest paths in ascending order. For each path, find a second disjoint path using Dijkstra algorithm such that the combined bandwidth of the two paths is at least X. If the length of the two found paths is less than the previous upper bound, set the current path length as an upper bound, and continue until all k-paths have been examined.

Loh et al [17] have addressed a more general version of the problem – finding multiple disjoint paths between source and destination nodes. They propose a polynomial-time heuristic algorithm, Maximum Bandwidth Algorithm (MBA), for solving this problem. The MBA algorithm creates two sets of edges. One set, called ES, contains edges ongoing from the source, and the other set, BS, contains all other edges. In both sets, the edges are sorted in descending order based on their bandwidth. At each round, take the highest-bandwidth edge in ES and remove all other edges in the network with lower bandwidths. Then, attempt to get the path with maximum bandwidth using Dijkstra algorithm. If no path is found, take the edges in BS in descending bandwidth, removing all edges from the network with lower-than-current bandwidth and run Dijkstra again. Before running Dijkstra algorithm, the cost of each edge is set to the result of subtracting the bandwidth of the link from a fixed number larger than the maximum bandwidth from the previous step. After finding the first maximum-bandwidth path, remove all the edges in that path from the network and continue to the next round with the next highest-bandwidth edge in ES. The authors have shown that the MBA algorithm produces the optimal disjoint paths in 99% of the cases using only 0.005% of the CPU time required using the optimal, but exponential, brute-force (BF) algorithm. Since it is one of the newest developed algorithms for the problem of concern, we have used the MBA algorithm for comparison with the algorithm proposed in this paper.

III. THE MODIFIED DIJKSTRA ALGORITHM

Sahni et al [10] have developed a modified version of Dijkstra algorithm to calculate the maximum bandwidth from a given source node s to a given destination node d. The algorithm shown in Figure 1 is based on Sahni's algorithm, but is extended to find the maximum bandwidth from a source node s to all other nodes in the network. The algorithm is also adapted to our syntax.

```
01  Algorithm MaxBandwidth(s)
02  for i = 1 to n do
03    if node[i] is a neighbor of s then
04      set maxbw of node[i] =
              bandwidth of link[s, i]
05    else
06      set maxbw of node[i] = 0
07    end if
08    set previous of node[i] = s
09  end for
10  set previous of node[s] = 0
11  label node[s] as PERMANENT
12
13  while exists node[i] with label = TENTATIVE do
14    find x = i with maxbw of node[i]
              not labeled as PERMANENT
15    if maxbw of node[x] = 0 then
16      exit #no more paths
17    else
18      label node[x] as PERMANENT
19    end if
20
21    for each neighbor node[v] of node[x] do
22      if  node[v] is not PERMANENT then
23        if minimum (maxbw of node[x],
              bandwidth of link [x,v])
              > maxbw of node[v]
24        then
25          set previous of node[v] = node[x]
26          set maxbw of node[v] =
                minimum(maxbw node[x],
                bandwidth of link[x, v])
27        end if
28      end if
29    end for
30  end while
```

Figure 1. The modified Dijkstra algorithm

The main differences between the modified algorithm and Dijkstra algorithm are explained. First, the search function in the modified algorithm searches for the maximum-bandwidth rather than the minimum-cost node (14-19). Second, the relaxation (neighbor update) phase chooses the largest between the bandwidth of the neighbor node and the minimum of the link bandwidth and the bandwidth of the current node (23-28). This is different from Dijkstra's relaxation phase, which chooses the minimum between the cost of the neighbor node and the sum of the link cost and the cost of the current node.

## IV. THE PROPOSED ALGORITHM

The algorithm presented in this paper can be summarized as follows. Work on finding two paths concurrently: one path is called R (red path) and the other is called B (blue path). At each instance of the algorithm execution, the algorithm finds the R path with maximum bandwidth, together with a node-disjoint path B with bandwidth not less than a specified limit. We will call the algorithm MLBDP, short for Max-Limit Bandwidth Disjoint Path.

Our algorithm uses the aforementioned modified Dijkstra algorithm to work on dual paths concurrently. To facilitate this approach, a virtual representation of the network is created with n×n virtual nodes (vnodes), where n is the number of nodes in the original network. Each virtual node is denoted by two symbols which represent the two current nodes in the dual path.

```
01  Algorithm MLBDP (s, limit)
02  #s is the source node in the original G
03  #limit is the minimum bandwidth allowed in
    the R path
04  #ss is source node in the virtual network
05  #n is the number of nodes in the original
    network G
06
07  for i = 1 to n do
08    for j = 1 to n do
09      visited list of vnode[ij] =
          new empty list
10      label vnode[ij] as TENTATIVE
11      set R.maxbw of vnode[ij] = 0
12      set B.maxbw of vnode[ij] = 0
13    end for
14  end for
15
16  for i = 1 to n do
17    label vnode[is] and vnode[si] as
        PERMANENT
18  end for
19
20  for i = 1 to n do
21    for j = 1 to n  do
22      if node[i] ≠ node[j] are neighbors of s
        and bandwidth of link[s,j] ≥ limit then
23        set R.maxbw of vnode[ij] =
                bandwidth of link [s,i]
24        set B.maxbw of vnode[ij] =
            bandwidth of link [s,j]
25        set previous of vnode[ij] =
            vnode[ss]
26      end if
27    end for
28  end for
29
30  #loop until reach all the destinations
    with two paths
31  while exists vnode[ii] with label =
    TENTATIVE do
32    #find vnode with maximum R.maxbw and
      make it PERMANENT
33    repeat
        find xy = ij with maximum R.maxbw of
34        vnode[ij] not labeled as PERMANENT
35      if R.maxbw of vnode[xy] = 0 then
36        exit #no more paths
37      else
38        label vnode[xy] as PERMANENT
39    until x ≠ y
40
41    for each neighbor node[v] of node[x] do
42      if  vnode[vy] is not PERMANENT then
43        if minimum (R.maxbw of vnode[xy],
            bandwidth of link [x,v]) >
            R.maxbw of vnode[vy]
44          and link[x, v] not in
              visited list of vnode[xy]
45        then
46          set previous of vnode[vy] =
              vnode[xy]
47          set visited list of vnode[vy] =
              visited list of vnode[xy]
              + link[v,x]
48          set R.maxbw of vnode[vy] =
              minimum(R.maxbw vnode[x,y],
              bandwidth of link[x,v])
49          set B.maxbw of vnode[vy] =
              B.maxbw of vnode[xy]
50        end if
51      end if
52    end for
53
54    for each neighbor node[u] of node[y] do
55      if  vnode[xu] is not PERMANENT  then
56        if R.maxbw of vnode[xy] ≥
            R.maxbw of vnode[xu]
57          and link[u, y] not in
              visited list of vnode[xy]
58          and minimum(B.maxbw of vnode[xy],
              bandwidth of link [y,u])
                ≥ limit
59        then
60          set previous of vnode[x,u] =
              vnode[xy]
61          set visited list of vnode[xu] =
```

```
                 visited list of vnode[xy]
                 + link[u,y]
62         R.maxbw of vnode[xu] =
              R.maxbw vnode[x,y]
63         B.maxbw of vnode[xu] =
              minmum( B.maxbw of vnode[x,y],
              bandwidth of link[y,u])
64       end if
65     end if
66   end for
67 end while
```

Figure 2.  The proposed MLBDP algorithm

The algorithm starts by initializing the R and B bandwidths of all virtual nodes to 0 and labeling them as TENTATIVE (7-14). The source virtual node [ss] which corresponds to the source, and any virtual node [si] or [is], are marked as PERMANENT (16-18).

Next, find all permutations of two nodes i and j which are neighbors of s such that the bandwidth of the link between the source and the second node j is at least equal to the limit. The maxbw of the R path (R.maxbw) is set to the bandwidth of link[s, i] and the maxbw of B path (B.maxbw) is set to the bandwidth of link[s,j] (20-28).

The main loop of the algorithm is next started (30-end), which remains until all virtual nodes [ii] are marked as PERMANENT. Note that reaching a virtual node with identical two indexes [ii] from the source means that the destination node with the corresponding single index [i] has been reached with two disjoint paths.

The repeat loop in (30-39) performs the search for R.maxbw vnode (vnode with max bandwidth R path) with maximum bandwidth greater than 0, if such vnode is found, it will be marked as PERMANENT. This is similar to the search phase in the modified Dijkstra algorithm. If the vnode [xy] has two identical constituents (x equals y), one destination is reached and so the algorithm doesn't examine the neighbor of that vnode, but continues to search for the next R.maxbw vnode. The loop will end when a vnode [xy] with x ≠ y is found. The vnode [xy] becomes the current working vnode.

The for loop in (41-52) iterates on all neighboring nodes [v] of the current working node [x] (first side of the path), as done in the relaxation phase of the modified Dijkstra algorithm. The minimum of the R.maxbw (current max bandwidth of R path) of the current node and the bandwidth of the link[x, v] is compared with the R.maxbw (current max bandwidth of R path) of virtual node [vy]. If this minimum is greater, this means a larger R.maxbw is found for the virtual node [vy]. Thus, the current R.maxbw of vnode [vy] is set to this minimum. The for loop in (54-66) iterates on all neighboring nodes [u] of the current working node [y] (second side of the path). It is similar to the previous for loop, except that the minimum of B.maxbw and the bandwidth of the link [y, u] is compared with the specified limit parameter, instead of the current max bandwidth of the B path.

In both loops it is important to ensure that the path is node-disjoint. This is accomplished by maintaining a visited list for each virtual node. This list contains nodes which have been traversed in the current pair of paths. The visited list is checked before a new node is added to the path (44, 57) and updated after the node is added to the path (47, 61).

Recall that the algorithm MLBDP finds the R path with maximum bandwidth, together with the B path with bandwidth ≥ limit. In order to find the node-disjoint path with the maximum total bandwidth, the algorithm MLBDP needs to be executed q times, where q is the number of unique link bandwidths. In each execution, the limit is set to one of the bandwidth values.

V. EXAMPLE OF THE PROPOSED ALGORITHM

To demonstrate how the MLBDP algorithm works, consider the network shown in Figure 3. It is required to find the maximum-bandwidth node-disjoint path from a to d. We will explain a single run of MLBDP with limit = 7.

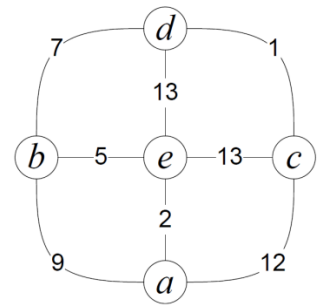

Figure 3.  Example network to explain the MLBDP algorithm.

The algorithm will start by initializing the source virtual node aa as PERMANENT and initializing the (R.maxbw, B.maxbw) values of the virtual nodes adjacent to aa, as follows: eb = (2, 9), ec = (2, 12), bc = (9, 12) and cb = (12, 9). Note that vnodes ce and be will not be initialized and will remain at (0, 0) because the bandwidth of the Blue-side link (B.maxbw) is less than the limit of 7. The vnode with largest R.maxbw is cb (12, 9) with R.maxbw = 12, so it will be chosen as current vnode and labeled as PERMANENT. From cb, the neighbor vnodes are eb (12, 9), db (1, 9) and cd (12, 7). Again, ce (12, 5) will not be considered because B.maxbw of ce = 5 < limit. Both eb (12, 9) and cd (12, 7) have largest R.maxbw, but eb has a larger B.maxbw so the eb will be chosen as current vnode and labeled as PERMANENT.

From eb, the neighbor vnodes are bb (5, 9), db (12, 9) and ed (12, 7). ee will not be considered because its B.maxbw = 5. db (12, 9) will be chosen as current and labeled as PERMANENT. Neighbors of db are dd (12, 7) and bb (7, 9). Note that de will not be considered for two reasons. First, node e has been traversed in the path from aa to db (aa-cb-eb-db). Thus, the visited list of vnode db currently has the nodes a, c and e. The second reason is because B.maxbw of de = 5 < limit. The current vnodes with highest R.maxbw are cd (12, 7), ed (12, 7) and dd (12, 7). They will be marked successively. When vnode dd is marked, this means that node d has been reached with the node-disjoint path (aa-cb-eb-db). The constituent paths are (a-c-e-d), (a-b-d) and the combined bandwidth is 12+7 = 19, which is the maximum node-disjoint path from a to d.

## VI. Performance Study

As mentioned in the Introduction, the problem of finding maximum-bandwidth disjoint path is considered NP-complete. As such, the optimal solution of the problem can only be obtained using ILP. Execution time of ILP is exponentially proportional to the number of nodes in the network but is guaranteed to find all possible disjoint paths. Heuristic solutions, on the other hand, require much less time, but may not always find the maximum-bandwidth disjoint paths. The purpose of this section is to compare the performance of the MLBDP algorithm presented in this paper with both ILP and the MBA algorithm mentioned in Section II. The comparison is based on both the execution time and the maximum-bandwidth disjoint paths found.

### A. ILP Formulation of the Problem

The ILP formulation of the problem of finding two disjoint paths has been developed in several related works. We use the formulation developed in [9]. Some modifications are made to the formulation to make it applicable for node-disjoint instead of link-disjoint paths.

Let V denote the set of nodes and E denote the set of links in the network. Links belonging to E are defined as pairs (u, v) that represent the nodes they are connecting. The source and destination nodes are denoted as s and t, respectively. The two paths are denoted as the red and blue paths. For each link (u, v) that belongs to E, four variables are defined: r(u, v), r(v, u), b(u, v) and b(v, u). These variables can take value 0 or 1. If the link from u to v belongs to the red path, r(u, v) = 1, else, r(u, v) = 0. Note that order of u of v is important, because it defines the direction of the path. Same can be said about b(u, v) and b(v, u). The bandwidths of the red and blue paths are denoted by yr and yb, respectively.

The function δ () is defined as follows:

$$\delta(x) = \begin{cases} 1 & x = s \\ -1 & x = t \\ 0 & \text{otherwise} \end{cases}$$

With the previous definitions, the ILP formulation can be stated as follows:

Maximize $y_r + y_b$ such that:

$$\sum_{v \in V} r(x,v) - \sum_{u \in V} r(u,x) = \delta(x), \quad \forall x \in V \quad (1)$$

$$\sum_{u \in V} b(x,u) - \sum_{u \in V} b(u,x) = \delta(x), \quad \forall x \in V \quad (2)$$

$$\sum_{v \in V} r(v,x) + b(v,x) = 2, \quad \text{for } x = t \quad (3)$$

$$\sum_{v \in V} r(v,x) + b(v,x) = 0, \quad \text{for } x = s \quad (4)$$

$$\sum_{v \in V} r(v,x) + b(v,x) \le 1, \forall x \in V, \quad x \notin \{s,t\} \quad (5)$$

$$y_r \le B_{uv} \cdot r(v,u) + M \cdot (1 - r(v,u)), \quad \forall (u,v) \in E \quad (6)$$

$$y_b \le B_{uv} \cdot b(v,u) + M \cdot (1 - b(v,u)), \quad \forall (u,v) \in E \quad (7)$$

$$y_r, y_b \ge 0 \quad (8)$$

$$r(u,v) + r(v,u) + b(u,v) + b(v,u) \le 1, \forall x(u,v) \in E \quad (9)$$

$$r(u,v), b(u,v) \in \{0,1\} \quad (10)$$

The requirement is to find the maximum total bandwidth of the red and blue paths, subject to the following conditions: Condition (1) ensures that the red path is connected. i.e., each node in the path is connected to two nodes, with the exception of s and t nodes, which have to be connected to only one node each. Condition (2) ensures the same for the blue path. Conditions (3), (4) and (5) ensure that, excluding the source and destination nodes, if any node is found in the red path, it is not in the blue path and vice versa. i.e., the path is node-disjoint. Conditions (6) and (7) ensure that bandwidth (yr and yb) of each path has the value of its lowest-bandwidth link. Condition (8) ensures that the bandwidth of each path is not less than zero. Condition (9) ensures that the two paths are also link-disjoint. Finally, condition (10) ensures that r(u, v) and b(u, v) can only take values 0, 1, which means that the link either does not or does exist in the path, respectively.

### B. Network Topologies and Setup

We study the performance of our MLBDP algorithm compared to ILP and the MBA heuristic algorithm developed in [17]. The comparative studies were performed on two network topologies: STC backbone network [18], shown in Figure 4, and ARPANET network [19], shown in Figure 5.

For both topologies, we examined networks with maximum possible bandwidth on any link equals to 10, 20, 50, 100, 200, 500, 1000, 2000 and 5000. For each case, we tested the maximum-bandwidth node-disjoint path obtained for each source and destination. The STC backbone network has 35 nodes and 45 links, while the ARPANET network has 20 nodes and 32 links.

The algorithms which have been compared include ILP, as formulated in this section, MBA algorithm, as described in Related Work, and our MLBDP algorithm. Recall that the MBA algorithm aims to find link-disjoint rather than node-disjoint maximum bandwidth paths. However, it can be easily modified to find node-disjoint paths. We made the necessary changes to make it node-disjoint in order to provide a fair comparison. We have used C# programming language for coding our MLBDP algorithm, as well as the MBA algorithm. ILP implementation was also programmed using C# with Microsoft Solver Foundation. Simulations were done on Core 2 Duo computers with about 2GHz speed.

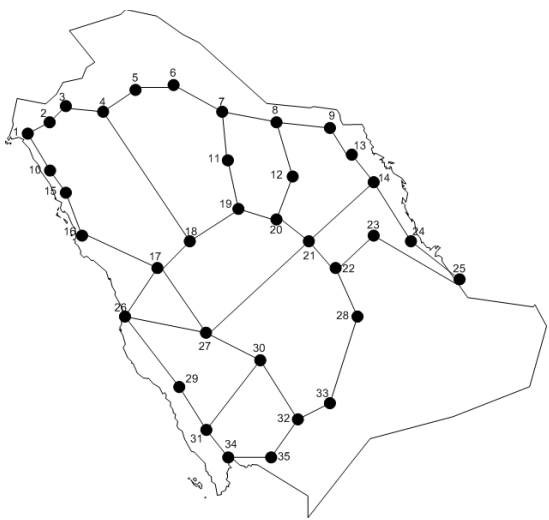

Figure 4. STC backbone network

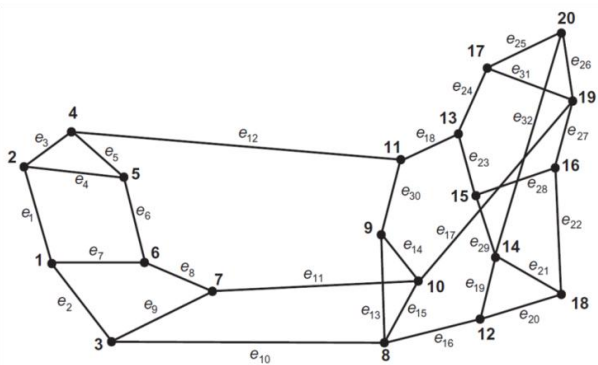

Figure 5. ARPANET network with 20 nodes

## C. Simulation Results

Simulation results for ARPANET network and STC backbone network are shown in Table I and Table II, respectively. Both networks have been tested under nine different sets of bandwidth capacities. The bandwidth of each link has been assigned a random value between 1 and a maximum value shown in the first column. Columns 2, 3 and 4 show the number of disjoint paths found, between all source-destination pairs, using MBA, our MLBDP algorithm, and ILP, respectively. For STC backbone network, there are 35 nodes. For each node taken as source, there can be at most 34 maximum-bandwidth node-disjoint paths, one for each destination. i.e., $35 \times 34 = 1190$ paths. For ARPANET network, this number is calculated as $20 \times 19 = 380$. We note that both networks are designed so that there is at least a node-disjoint path between each pair of nodes. Columns 5, 6, and 7 show the execution time in milliseconds it took each of the three algorithms to find the maximum-bandwidth node-disjoint paths. It should be noted that the solutions obtained using ILP are considered optimal and can be used to benchmark other algorithms. The maximum bandwidth obtained by the MBA algorithm in each case has been subtracted from the maximum bandwidth obtained by ILP for the same case. The total difference (sum of subtraction results) is shown in column 8 and the average difference (average of subtraction results) is shown in column 9. Columns 10 and 11 show the total and average differences calculated in the same manner between our MLBDP algorithm and ILP.

From the results, it can be observed that ILP execution time is much higher than MBA and MLBDP algorithms. Note that a log scale has been used to make clearer representation in Figure 7 and Figure 9. Although our MLBDP algorithm takes longer time to finish than MBA algorithm, it was able to find significantly higher number of disjoint paths than MBA and thus was able to find higher maximum-bandwidth paths. It can be seen from Figure 6 and Figure 8 that our MLBDP algorithm was able to find all possible disjoint paths obtained by ILP in a much lower execution time (about two orders of magnitude less than ILP). By looking at columns 8 to 11 in Table I and Table II, it can be observed that there is a difference between the maximum bandwidth obtained by MBA and by ILP. The bandwidth difference is higher at lower rows because these rows have higher deviation between the bandwidths available in each link. The reason of MBA failure is the two-step approach used in it. i.e., after finding the first maximum-bandwidth path, all links in this path are removed from the network. The removed links can at worst prevent an existing disjoint path from being found, or at best eliminate a path with higher combined maximum bandwidth. On the other hand, it can be seen that our MLBDP algorithm has a zero difference with ILP in the maximum bandwidth found. That's because the MLBDP algorithm works concurrently on the two disjoint paths, minimizing the probability of missing a better candidate path.

Graphical representations of the results are shown in Figure 6 and Figure 7 for STC backbone network. For ARPANET network, the results are shown in Figure 8 and Figure 9.

TABLE I. SIMULATION RESULTS FOR STC BACKBONE NETWORK

| Max BW | Number of Disjoint Paths Found | | | Execution Time | | | Difference ILP-MBA | | Difference ILP-MLBDP | |
|---|---|---|---|---|---|---|---|---|---|---|
| | *MBA* | *MLBDP* | *ILP* | *MBA* | *MLBDP* | *ILP* | *Total* | *Average* | *Total* | *Average* |
| 10 | 988 | 1190 | 1190 | 1827.1 | 4103.2 | 138872.9 | 334 | 0.28 | 0 | 0 |
| 20 | 934 | 1190 | 1190 | 2354.1 | 6869.3 | 173326.9 | 566 | 0.48 | 0 | 0 |
| 50 | 930 | 1190 | 1190 | 3028 | 12319.7 | 205795.7 | 1124 | 0.94 | 0 | 0 |
| 100 | 930 | 1190 | 1190 | 3146.1 | 14305.8 | 202170.5 | 1938 | 1.63 | 0 | 0 |
| 200 | 929 | 1190 | 1190 | 3161.1 | 16949.9 | 192363 | 3601 | 3.03 | 0 | 0 |
| 500 | 929 | 1190 | 1190 | 3390.1 | 16927.9 | 193892 | 8522 | 7.16 | 0 | 0 |
| 1000 | 929 | 1190 | 1190 | 3427.1 | 18523 | 231039.2 | 16726 | 14.06 | 0 | 0 |
| 2000 | 929 | 1190 | 1190 | 3603.2 | 18923 | 220709.6 | 33194 | 27.89 | 0 | 0 |
| 5000 | 897 | 1190 | 1190 | 3571.2 | 18646 | 233137.3 | 131453 | 110.46 | 0 | 0 |

TABLE II. SIMULATION RESULTS FOR ARPANET NETWORK

| Max BW | Number of Disjoint Paths Found | | | Execution Time | | | Difference ILP-MBA | | Difference ILP-MLBDP | |
|---|---|---|---|---|---|---|---|---|---|---|
| | *MBA* | *MLBDP* | *ILP* | *MBA* | *MLBDP* | *ILP* | *Total* | *Average* | *Total* | *Average* |
| 10 | 362 | 380 | 380 | 436 | 429 | 89235.1 | 82 | 0.22 | 0 | 0 |
| 20 | 362 | 380 | 380 | 484 | 888 | 116026 | 103 | 0.27 | 0 | 0 |
| 50 | 362 | 380 | 380 | 568 | 1258 | 100572 | 255 | 0.67 | 0 | 0 |
| 100 | 362 | 380 | 380 | 557 | 1158 | 118052 | 468 | 1.23 | 0 | 0 |
| 200 | 357 | 380 | 380 | 588 | 1352 | 118377 | 2278 | 5.99 | 0 | 0 |
| 500 | 357 | 380 | 380 | 728 | 1417 | 159979 | 5811 | 15.29 | 0 | 0 |
| 1000 | 357 | 380 | 380 | 680 | 1531 | 138376 | 11572 | 30.45 | 0 | 0 |
| 2000 | 357 | 380 | 380 | 656 | 1420 | 136474 | 22969 | 60.44 | 0 | 0 |
| 5000 | 357 | 380 | 380 | 566 | 1490 | 131096 | 57262 | 150.69 | 0 | 0 |

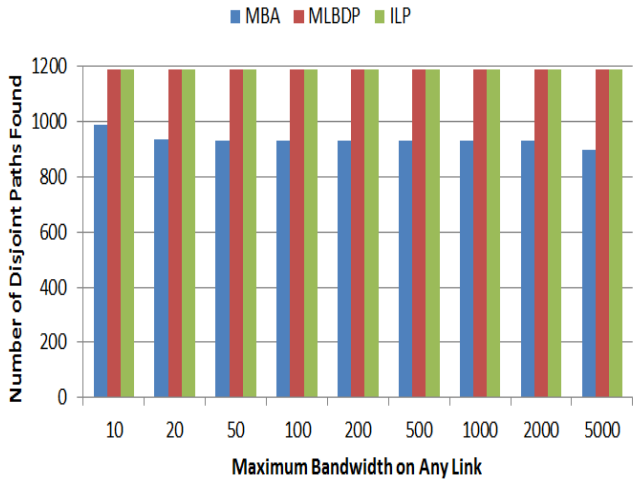

Figure 6. Number of disjoint paths found - STC backbone network

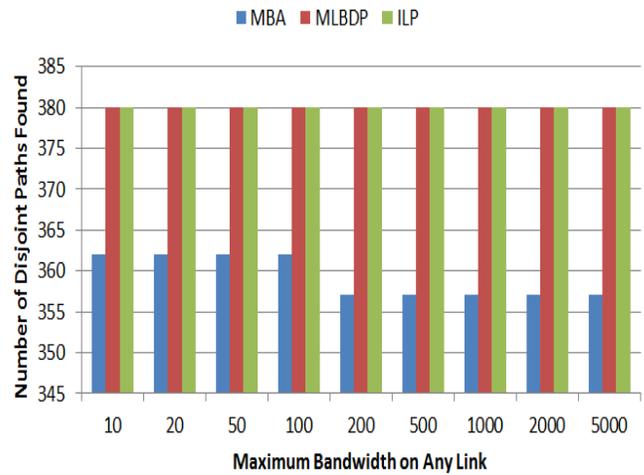

Figure 8. Number of disjoint paths found - ARPANET network

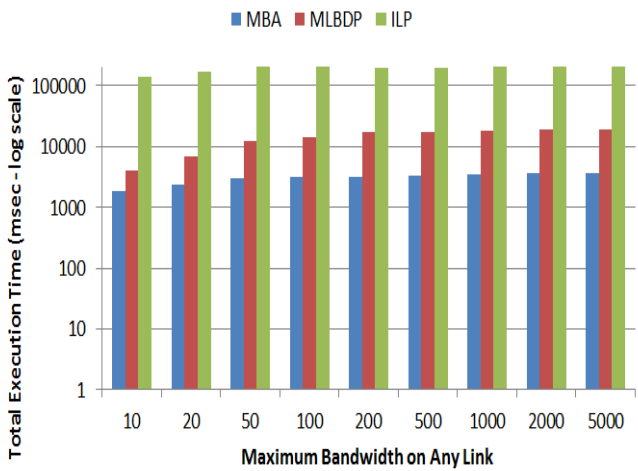

Figure 7. Total execution time - STC backbone network

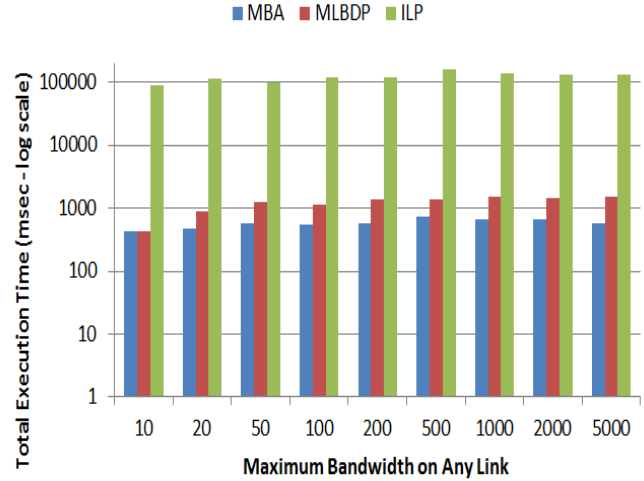

Figure 9. Total execution time - ARPANET network

## VII. ANALYTICAL STUDY

From the numerical results in the previous section, we can observe that our MLBDP algorithm finishes in polynomial time. The goal of this section is to provide a general estimation of the execution time.

It has been shown in [20] that Dijkstra algorithm works in:

$$\text{Run}(\text{Dijkstra}) = O(m + n \log n) \quad (11)$$

where n is the number of nodes and m is the number of links. The MLBDP algorithm uses a virtual topology with n×n virtual nodes and with 2nm virtual links. To see that the number of virtual links is 2mn, it can be seen from the algorithm description in Figure 2 that the number of links originating from each virtual node (vnode) equals the sum of the number of links originating from both of its constituent (actual) nodes. Let the number of links coming from nodes i and j equal ki and kj, respectively. The number of links originating from vnode[ij] equals ki+kj.

To calculate the total number of links K in the virtual topology, note that each link is connected to two nodes. If we add the number of links originating from each node, every link will be counted twice. Thus, the summation of the number of links should be divided by 2., i.e.

$$K = \frac{1}{2}\sum_{i=1}^{n}\sum_{j=1}^{n}(k_i + k_j) = \frac{1}{2}\sum_{i=1}^{n}\sum_{j=1}^{n}k_i + \frac{1}{2}\sum_{i=1}^{n}\sum_{j=1}^{n}k_j \quad (12)$$

We can see that $\sum_{j=1}^{n}k_j = 2m$. Similarly $\sum_{i=1}^{n}k_i = 2m$. Substituting in (12) yields:

$$\frac{1}{2}\sum_{i=1}^{n}n \cdot k_i + \frac{1}{2}\sum_{i=1}^{n}2m = \left(\frac{n}{2}\right)(2m) + (2m)\left(\frac{1}{2}\right)(n) = 2mn \quad (13)$$

From (11) and (13), a single run of the MLBDP algorithm yields a run time of approximately:

$$\text{Run}(\text{MLBDP}) \approx O(2mn + n^2 \log n^2)$$
$$\approx O(2mn + 2n^2 \log n) \quad (14)$$

Recall from Section 4 that the MLBDP algorithm needs to be executed q times, where q is the number of unique link bandwidths. In the worst case, where each of the m links in the network has a unique bandwidth, q = m. Thus, the worst-case full run of MLBDP algorithm yields a run time of approximately:

$$\text{FullRun}(\text{MLBDP}) \approx O(m(2mn + 2n^2 \log n))$$
$$\approx O(2m^2 n + 2mn^2 \log n) \quad (15)$$

## VIII. SUMMARY AND CONCLUSIONS

This paper has presented a new algorithm, Max-Limit Bandwidth Disjoint Path (MLBDP), for finding node-disjoint paths with maximum combined bandwidth. The algorithm works on multiple iterations. At each iteration, the algorithm works on finding two paths concurrently: one with maximum bandwidth and another with bandwidth greater than a certain limit. The limit takes the values of all possible unique link bandwidths. The presented algorithm uses a modified Dijkstra algorithm and a virtual network topology. A performance comparison has been done for the MLBDP algorithm to a modern heuristic, MBA algorithm, and to the optimal solution using Integer Linear Programming (ILP). The simulation studies have shown that the MLBDP algorithm was able to obtain results identical to the ILP solution at a significantly lower execution time. In addition, the MLBDP avoids the MBA problem of missing valid disjoint paths because it works on the two disjoin paths concurrently. Thus, despite the slightly additional execution time, the MLBDP algorithm offers an overall better performance over the MBA algorithm. Future enhancements of the presented MLBDP algorithm can be done to reduce the execution time. Also, the algorithm can be modified to find maximum-bandwidth link-disjoint paths.

ACKNOWLEDGMENT

The author would like to thank the Research Center of the College of Computer and Information Sciences, King Saud University, for their support.

AUTHORS PROFILE

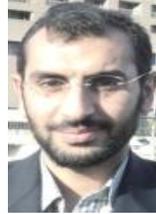

**Mostafa H. Dahshan** has received his B.S. degree in Computer Engineering from Cairo University, Egypt in 1999. He received his M.S. in Telecomm. Systems and Ph.D. in Electrical and Computer Engineering from the University of Oklahoma, USA in 2002 and 2006, respectively. He is currently an Assistant Professor of Computer Engineering at the College of Computer and Information Sciences, King Saud University, Saudi Arabia. His current research interests include Network Protocols, Performance, Reliability and Security.